\documentclass[preprint,12pt,5p,times,twocolumn]{elsarticle}
\usepackage{epsfig}

\journal{Sol. Energy Mater. Sol. Cells}
\begin{document}
\begin{frontmatter}
\title{Photovoltaic performance of n-type SnS active layer in ITO/PEDOT:PSS/SnS/Al structure}

\author[doe,khalsa]{Priyal Jain}
\author[khalsa]{P.Arun \corref{cor1}\fnref{fn1}}

\cortext[cor1]{Corresponding author}
\fntext[fn1]{(T) +91 11 29258401 (Email) arunp92@physics.du.ac.in}
\address[doe]{Department of Electronic Science, University of Delhi, 
South Campus, Delhi 110021, INDIA}
\address[khalsa]{Material Science Research Lab, S.G.T.B. Khalsa College,
University of Delhi, Delhi - 110 007, INDIA}

\begin{abstract}
The present paper discusses the influence of Tin Sulphide's grain size on the 
performance of ITO/PEDOT:PSS/SnS/Al structured solar cells fabricated by 
thermal evaporation. 
The grain sizes were maintained in the range of 11-18~nm by controlling the 
thickness of SnS films. While the open circuit voltage ($V_{oc}$) was found 
to be a constant for this structure, parameters such as short 
circuit current density ($J_{sc}$), series resistance ($R_s$), parallel 
resistance ($R_p$), ideality factor and the overall efficiency were found to 
be dependent on the SnS grain size and incident light intensity. The 
experimental work directly reconfirms the theoretical results and ideas 
raised in literature by early researchers.

\end{abstract}
\begin{keyword}
Thin Films; Chalcogenides; Solar Cells
\end{keyword}

\end{frontmatter}
\section{Introduction}
With the depleting reserves of fossil fuels and its adverse 
effects on the environment, efforts are being made to search for 
environment friendly and renewable sources of energy. Solar energy seems to 
be the most promising candidate. These days research is focussed on the 
development and fabrication of high efficiency and low cost photo-voltaic 
devices. Photo-voltaics are usually p-n devices with active region having 
enough absorption coefficient to absorb the incident light yet thin enough
for charge carriers to reach their respective electrodes without
recombination. Thin film technology suits the demands and active layer of 
few micrometer thickness are fabricated by various fabrication techniques.

Inorganic thin film solar cells have always been considered as one of the 
main cost-effective options for the future photovoltaic devices. In recent 
years chalcogenide materials such as PbS~\cite{pbs}, PbSe~\cite{pbse}, 
CdS~\cite{cds}, CdTe~\cite{cdte}, SnS~\cite{ristov} and ${\rm Sn_xPb_{1-x}S}$ 
\cite{mix} have attracted attention. The best efficiency that has been 
reported was using CdS/CdTe at ${\rm \sim 16.5\%}$~\cite{record}. However, 
due to the toxic nature of Cadmium and its compounds, Cadmium compounds rule 
themselves out as a candidate~\cite{reddy}. Tin Sulphide (SnS) promises to be 
the most suitable candidate for solar cells~\cite{ristov,prasert} due to its 
optimum optical 
properties which are tunable by varying the grain size and/ or lattice 
parameters~\cite{pj1, pj2, pj3}. SnS films also have high absorption 
coefficients~\cite{thangaraju, noguchi} and optical band-gap in the range of 
1.1-2.1~eV~\cite{gao,sohila}. On top of all this, SnS is non toxic~\cite{zhi} 
and easily available~\cite{yue}. However, structures with only SnS do not show 
photo-voltaic behavior~\cite{alex}. 
\begin{table*}[t]

{\bf Table 1:} {\sl Comparison of properties of SnS based solar 
cells from the literature listed in order of increasing efficiency.}
\begin{center}
\begin{tabular}{c c c c c c c}
\hline 
  S.No. & Structure &  Grain Size (nm) & $J_{sc}$~(mA/$cm^2$) & $V_{oc}$ (V) &
${\rm \eta ~(\%)}$ & Ref\\ \hline
  1.   &ITO/PEDOT:PSS/Polymer:SnS/Al & 10 & 0.055 & 0.84 & 0.0205 & \cite{zhi}\\
  2.   &ITO/CdS/SnS/Al & 20-40 & - & 0.2 & 0.09 & \cite{ghosh}\\
  3.   &ITO/CdS/SnS/Al & - & 1.23 & 0.37 & 0.2 & \cite{david}\\
  4.   &FTO/${\rm TiO_{2}}$/SnS Quantum dots/Pt & 5.06 & 2.3 & 0.504 & 0.2 & \cite{deepa}\\  
  5.   &ITO/PbS/SnS/Al & 9.6 & 1.2 & 0.44 & 0.27 & \cite{alex} \\ 
  6.   &ITO/CdS/SnS/Ag  & 300  & 7 & 0.12  & 0.29 & \cite{noguchi}  \\   
  7.   &${\rm SnO_2}$/SnS/CdS:In/In & 350 & 9.6 & 0.26 & 1.3 & \cite{reddy}\\  \hline
\end{tabular}
\end{center} 
\end{table*}

Research work have hence shifted focus towards hetero-junction structures of 
SnS, for example CdS/SnS~\cite{ghosh,david}, SnS/PbS~\cite{alex}, SnS quantum 
dots on ${\rm TiO_2}$~\cite{deepa} and SnS nanoparticles embedded in 
polymer~\cite{zhi} hetero-junction structures. Table~1 summarizes the 
photo-voltaic performance along with grain size of the SnS active layer. 
Interestingly, Noguchi et al~\cite{noguchi} obtained a more efficient 
photo-voltaic (${\rm \eta \sim 0.29\%}$) with the same structure used by 
Ghosh et al~\cite{ghosh} and David et al~\cite{david}. While David et 
al~\cite{david} 
obtained their active SnS layer by chemical deposition, Noguchi 
et al\cite{noguchi} used thermal evaporation to obtain their active layer. 
Interesting, for the same structure, Reddy et al.~\cite{reddy} improved upon 
Noguchi et al~\cite{noguchi} using SnS films of larger grain size obtained by 
spray pyrolysis (${\rm \eta \sim 1.30\%}$). These studies show that there is a 
co-relation between the active layer's grain size and structure's efficiency.

It is clear from the above discussion and results listed in Table~1 that the 
granularity of the film affects the performance of photo-voltaic devices.
Hence, it would also be of fundamental interest to enquire how the grain size 
influences solar cell parameters such as open circuit voltage (${\rm
V_{oc}}$), short circuit current density (${\rm J_{sc}}$) etc.
Therefore, we decided to revisit a simple photo-voltaic structure involving 
a tin sulphide active layer. The effect of the active layer's grain size 
and illumination intensity
on the device parameters (and hence its performance) have been investigated.


\section{Experimental Details}

\begin{figure}[h!]
\begin{center}
\epsfig{file=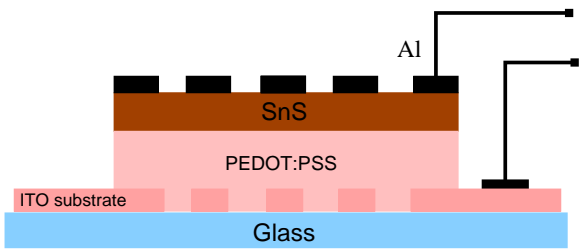, width=3.0in}
\hskip 2cm 
\label{cell}
\caption{\sl Figure shows the structure of our solar cell with layers of 
ITO/PEDOT:PSS/SnS/Al.}
\end{center}
\end{figure}
For the present study, Indium Tin Oxide (ITO) substrates were used. 
ITO has been a popular choice as a substrate in solar cells
because of its high conductivity and transparency at room 
temperature~\cite{gunes}. This
feature allows the substrate to double as an electrode of the device on
proper etching. We have used an ITO substrate of low resistivity 
(${\rm 10-15~\Omega cm}$). On the etched ITO substrates, a thin buffer layer 
(200~nm) of solar grade PEDOT:PSS (1.3~weight~\%) was spin coated, followed by
which the substrates were dried in a vacuum furnace at 373~K. This gives an 
uniform, conducting and transparent layer that stabilizes the etched surface of 
ITO substrate. This layer also favors the hole injection and collection at the 
electrode~\cite{kim}. Our attempts to fabricate simple ITO/SnS/Al structures 
(i.e.
without PEDOT:PSS) failed to give typical IV characteristics of
photo-voltaic devices. The failure of the simple ITO/SnS/Al structures has been 
reported earlier too~\cite{alex}. The active layer (SnS thin film) was 
deposited on the ITO/PEDOT:PSS substrate which was maintained at 
room temperature. During the deposition, glass substrates were placed
alongside ITO/PEDOT:PSS substrates. Films on glass substrates were used only 
for hotprobe tests to determine the nature of charge carriers present in them. 
The films grown on ITO/PEDOT:PSS were divided into two parts, one for solar
cell fabrication and second for characterization. The photovoltaic structure 
was completed by depositing Aluminium electrodes, again by thermal evaporation
using standard mask giving electrodes of area ${\rm \sim 0.16cm^2}$. Thus, we 
zeroed down to an ITO/PEDOT:PSS/SnS/Al structure 
similar to that used by Wang et al~\cite{zhi}. Fig~1 shows the cross-sectional 
view of the final structure. The current-voltage (J-V) measurements were done 
with a computer monitored Keithley 2400 source meter unit. A solar simulator 
of variable illumination intensity served as the light source. 
\begin{figure}[t!!!]
\epsfig{file=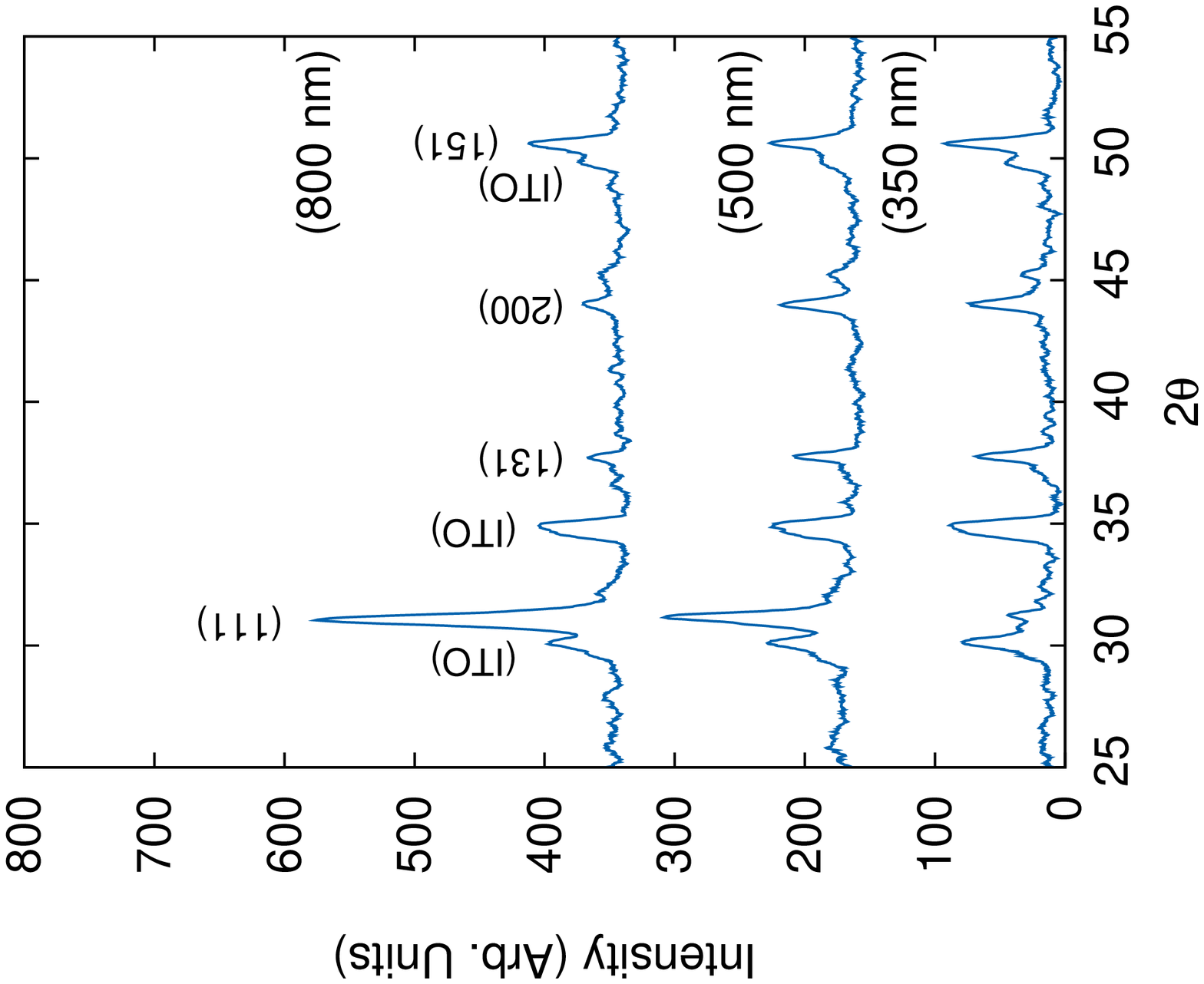, width=3.5in, angle=-90}
\vskip 1.25cm
\epsfig{file=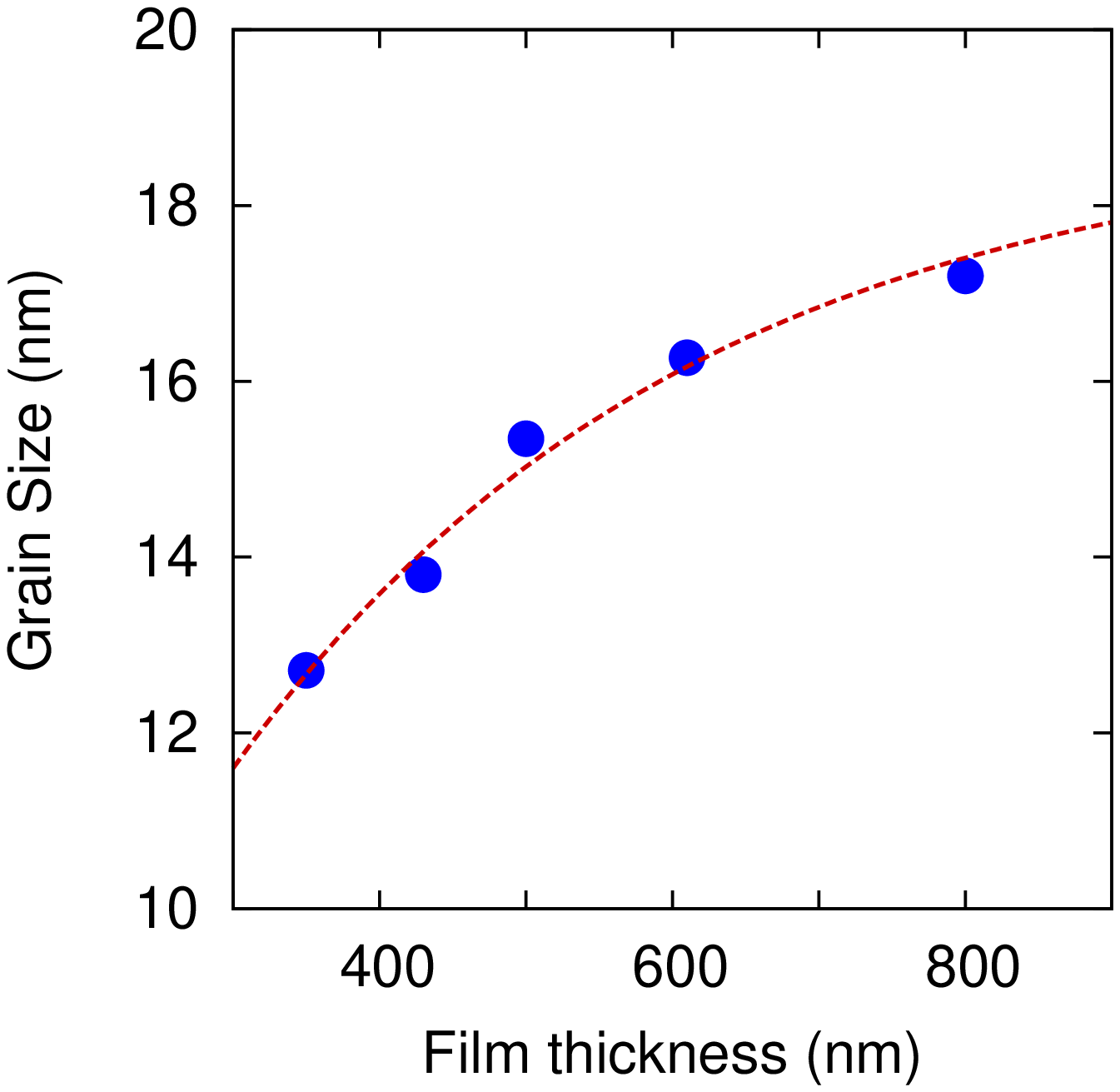, width=2.9in, angle=-90}
\label{xrd1}
\caption{\sl (A) X-Ray Diffraction Profiles of selected SnS films grown on 
ITO glass substrates. Sample identification with their Miller indices are 
indicated. Fig~(B) shows the variation of grain size with film
thickness.}
\end{figure}

For characterization the thicknesses of the active layers were measured using 
Veeco's Dektak Surface Profiler (150). The nano/poly-crystallinity of the 
active layers was determined by X-ray diffraction (XRD) (Bruker D8 X-ray 
Diffractometer) of ${\rm CuK\alpha}$ radiation (${\lambda \sim 1.5406~\AA}$) 
in the ${\rm \theta-2\theta}$ configuration. The surface morphology of the 
samples was determined using a Field Emission-Scanning Electron Microscope 
(FE-SEM FEI-Quanta 200F) at an accelerating voltage of 
10~kV. 

\begin{figure}[h!!!]
\begin{center}
\epsfig{file=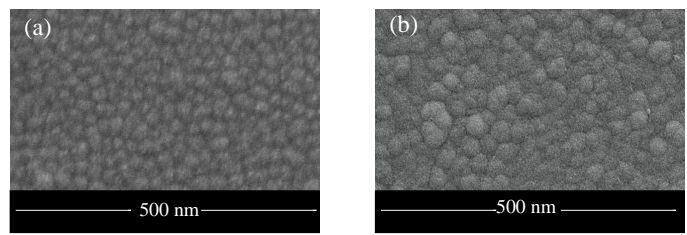, width=3.50in}
\end{center}
\label{sem}
\vskip -0.5cm
\caption{\sl SEM micrographs of SnS films (a) of 500~nm and (b) 800~nm
thickness.}
\end{figure}
\section {Characterization of SnS films} 
\begin{figure}[b!!!]
\begin{center}
\epsfig{file=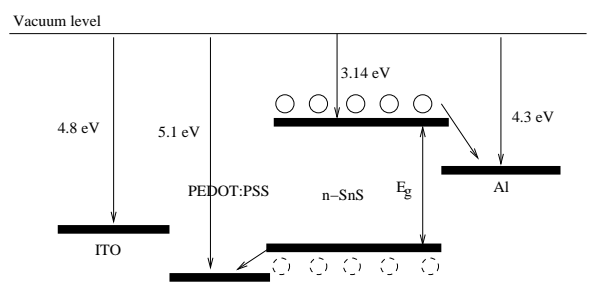, width=3.0in}
\hskip 2cm 
\label{energy}
\caption{\sl Energy level diagram of the ITO/PEDOT:PSS/SnS/Al structure.}
\end{center}
\end{figure}
\begin{figure*}[t!!!]
\begin{center}
\epsfig{file=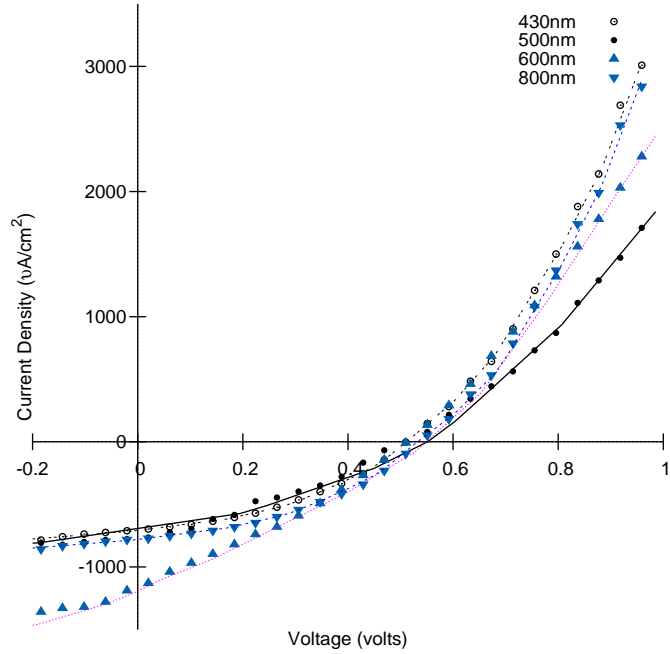, width=3.5in, angle=-90}
\end{center}
\label{jv}
\caption{\sl J-V characteristics of photovoltaic structures 
ITO/PEDOT:PSS/SnS/Al. Curves (i)-(iv) show the characteristics for different 
thicknesses of SnS obtained under illumination (${\rm 100mW/cm^2}$).}
\end{figure*}
The structure of the SnS films on ITO/PEDOT:PSS substrates were studied using 
X-ray 
Diffraction. Fig~2 shows the diffraction profiles of the SnS films of various 
thicknesses. Four prominent diffraction peaks at ${\rm 2\theta 
\approx 31^o}$, ${\rm 38^o}$, ${\rm 43^o}$ and ${\rm 51^o}$ were identified 
from the (111), (131), (200) and (151) crystal planes respectively of tin
sulphide with orthorhombic unit cell structure of lattice parameters 
a=4.148$\AA$, b=11.36$\AA$ and c=4.177$\AA$ (ASTM No.83-1758). The remaining
peaks of fig~2 (marked) are those of the ITO layer~\cite{biswajit}. 
The SnS films on ITO substrates were oriented with the the (111) planes 
perpendicular to the substrate~\cite {biswajit, devika}. The average grain 
size of the films were also 
calculated using the well known Scherrer formula~\cite{cullity}. The grain 
size was found to increase with film thickness (fig~2). The grain size
tends to saturate for large thicknesses (${\rm >700~nm}$). However, the grain size 
variation with thickness is linear for the range 350-600~nm.

Fig~3 shows the 
Scanning Electron Microscope (SEM) images of the SnS films (thickness 500 and 
800~nm). The images show that the grains are spherical in nature. In our
recent works \cite{pj1,pj2, pj3}, we have grown SnS films on glass substrates 
which resulted in films oriented with (040) planes parallel to the
substrate or in other words the larger `b' edge of the unit cell were
arranged along the substrate. This resulted in elongated SnS grains on glass
substrates. However, the (111) orientation on ITO substrates seems to prefer
spherical grain formation~\cite{biswajit,devika,saj}. Hotprobe measurements 
reveal that the SnS films used in this study are n-type without exception. 
Literature reports SnS films can either be of p-type or n-type, depending on 
fabrication conditions~\cite{saj}. Clearly, we have been able to maintain our 
sample fabrication conditions throughout our study. Also, considering that 
our SnS films were n-type, PEDOT:PSS conducting polymer serves as the p-type 
layer, thus forming an inorganic-organic hetergenous solar cell 
structure~\cite{nanoscale}. Fig~4 shows the energy level diagram of our structure.
The value of work functions were take from literature
\cite{nanoscale,devika2}.
The highest occupied molecular orbital (HOMO) level of the PEDOT:PSS lies 
below the valence band of SnS, hence the photo-generated holes at the 
junction easily drift towards the PEDOT:PSS layer while it acts as a barrier 
for electrons generated. The electrons move towards the Al electrode.

Summarizing our above results, tin sulphide films grown by thermal
evaporation give oriented films of n-type on ITO substrates, whose spherical 
shaped 
grains increase in size with increasing film thickness. We believe the above 
results would be of essence in establishing a relation between performance
and fabrication conditions.

\section{Solar Cells}
For characterization of our solar cells we have made Current-Voltage 
(I-V) measurements for all our samples. It is an industry standard to make
measurements of solar-cells at room temperature 
with illumination intensity of 100~mW/${cm^2}$ of Air Mass (AM) 1.5
spectrum for performance comparison. While we have made measurements at 
various illumination
intensities (summarized in Table~2), in the following passages we detail our
device's performance under ${\rm 100~mW/cm^2}$ illumination.

Fig~5 gives the J-V characteristics of our solar cells. 
Various models like single-diode, two-diode models~\cite{model} etc. are used 
to evaluate
the characterizing parameters of solar cells from which their efficiencies
are calculated. While the single-diode model emphasizes on the junction 
formed
due to the `p' and `n' layer, the two-diode model takes into account the
recombinations occuring across the grain boundaries of a polycrystalline 
active layer. If the recombination across the grain boundaries are 
insignificant compared to the contributions from the pn junction or if its 
contribution is 
constant (i.e. is not a function of applied voltage/current) then the single 
diode model can be applied on polycrystalline active layers too~\cite{model}. 
Decision is based on the nature of J-V
curve in the first quadrant. For linear variation~\cite{model, sze} a 
single-diode 
best models the behavior of the solar cells. Fig~6 shows the equivalent 
single-diode model of a solar cell. All the research
groups listed in Table~1  agree that nano/polycrystalline SnS has linear J-V 
characteristics in the first quadrant and hence have used the single diode 
model to characterize their solar cells. However, it is interesting to note
that while the recombination across grain boundary might be insignificant, 
grain size does play a significant role as can be seen from improved 
efficiency with grain size (Table~1). 

\begin{figure}[h!!]
\begin{center}
\epsfig{file=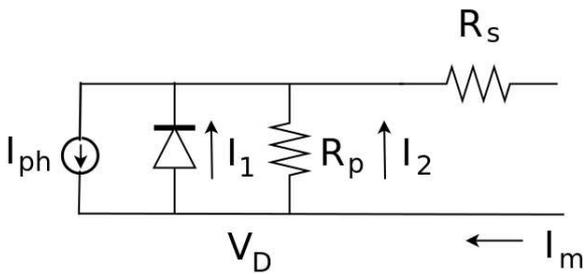, width=3.0in}
\end{center}
\vskip -0.5cm
\label{diode}
\caption{\sl Single diode equivalent circuit of a solar cell. }
\end{figure}

As can be understood from fig~6, we need to determine the series resistance 
(${\rm R_s}$) and parallel resistance (${\rm R_p}$) for characterizing the 
solar cell completely. This is important since these internal resistances 
decide parameters like the open circuit voltage (${\rm V_{oc}}$), the short 
circuit current density (${\rm J_{sc}}$), the ideality factor (n), the Fill 
Factor (FF) and ultimately the peak power (${\rm
P_{max}}$). The internal resistances were estimated from the slopes of the
J-V curve as per the single diode model \cite{zhi} 
\begin{eqnarray}
R_s &=& \left({dv \over di}\right)_{v=V_{oc}}\\
R_p &=& \left({dv\over di}\right)_{v=0}
\end{eqnarray}
\begin{figure}[h!!!]
\begin{center}
\epsfig{file=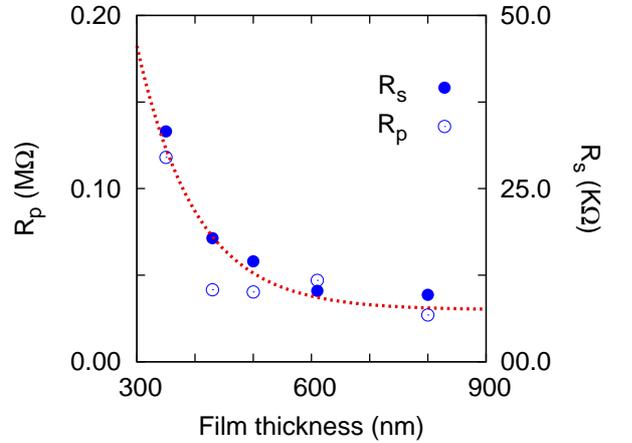, width=2.7in, angle=-90}
\end{center}
\label{rsvsgs}
\vskip -0.5cm
\caption{\sl Series resistance (${\rm R_s}$) and Parallel Resistance (${\rm
R_p}$) of solar cells as a function of the SnS active layers thickness.} 
\end{figure}
${\rm V_{oc}}$ is the open circuit voltage i.e. the voltage at which no 
current flows through the external circuit. This is directly obtained from 
fig~5 as the point where experimental curve cuts the `X'-axis. The short 
circuit current (${\rm I_{sc}}$), is the current that flows in the external 
circuit when the electrodes are short circuited (again, obtained from 
fig~5 as the point where experimental curve cuts the `Y'-axis). We have 
evaluated ${\rm R_s}$ and ${\rm R_p}$ as a function of film thickness. 
Both the resistances are found to decrease with increasing film thickness 
(fig~7). Interestingly, the trends are identical for both the resistances 
with ${\rm R_p > R_s}$, (${\rm \approx R_p=4 \times R_s}$). The variation in
these resistances with grain size also shows a similar trend (not shown). The 
resistance ${\rm R_s}$ appears from
the Al electrode to an imaginary point within the bulk of the SnS active
layer. Thinner films have smaller grain size and in-turn larger inter-grain 
boundary voids. This can be seen in the micrographs of fig~3. These
inter-grain voids would contribute large resistances which decreases with
increasing film thickness \cite{parun}.
Fig~8a shows the variation in the open circuit voltage and short circuit 
current density with the grain size. The open circuit voltage was found to
be constant (${\rm \approx 0.51~V}$) within experimental error. The value of 
${\rm V_{oc}}$ we report is higher as compared to similar structures reported 
in literature~\cite{reddy,noguchi}. The open circuit voltage is essentially 
the voltage across ${\rm R_p}$ and is
reported to be constant for large values of ${\rm R_p}$~\cite{renu}.
Considering Renu et al~\cite{renu} have indicated ${\rm R_p >1~K\Omega}$ 
as large, with our parallel resistances laying in the range of ${\rm
27-118~K\Omega}$, we expect ${\rm V_{oc}}$ to be constant. 
\begin{figure}
\begin{center}
\epsfig{file=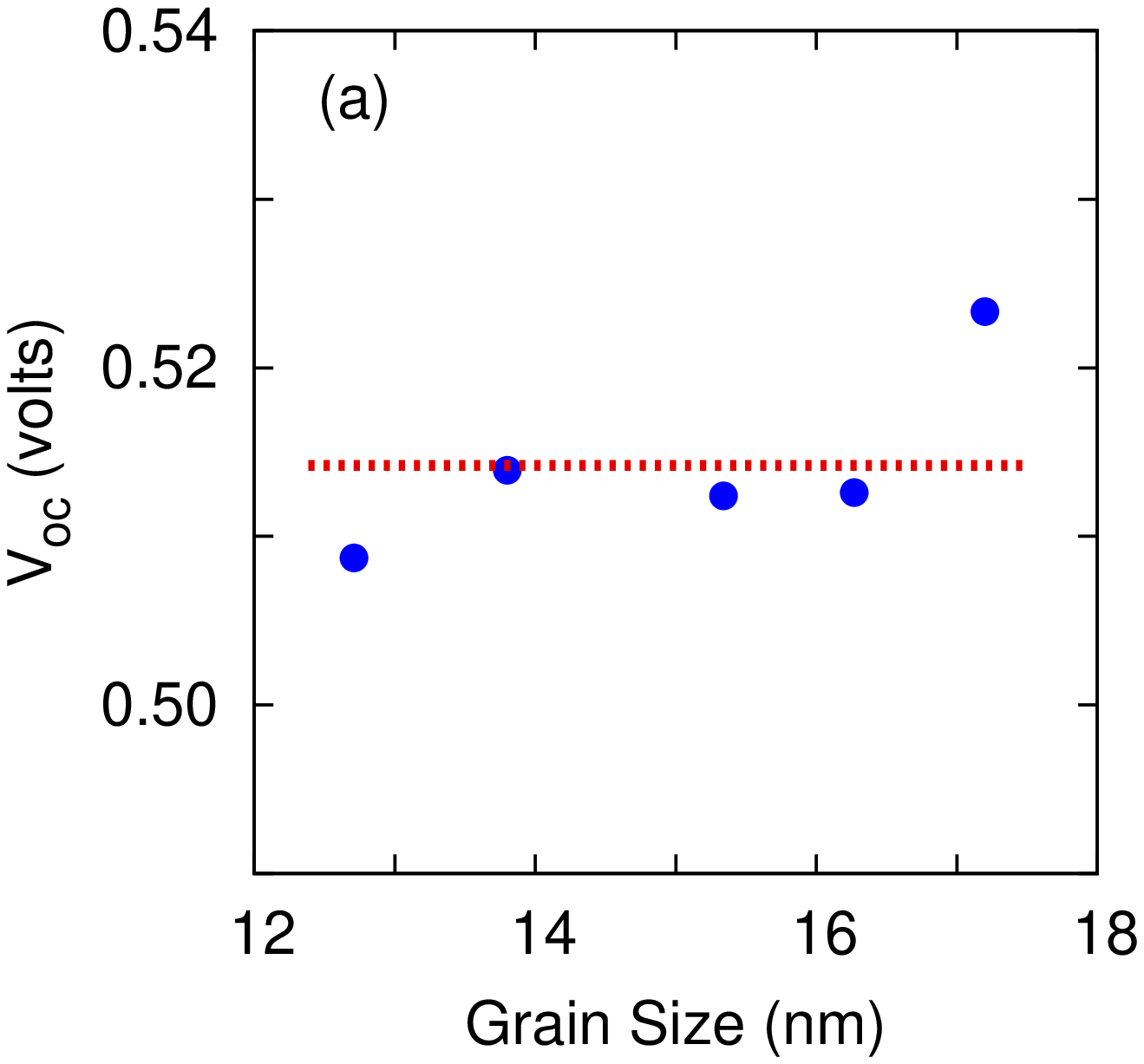, width=2.7in, angle=-90}
\vskip -0.5cm
\epsfig{file=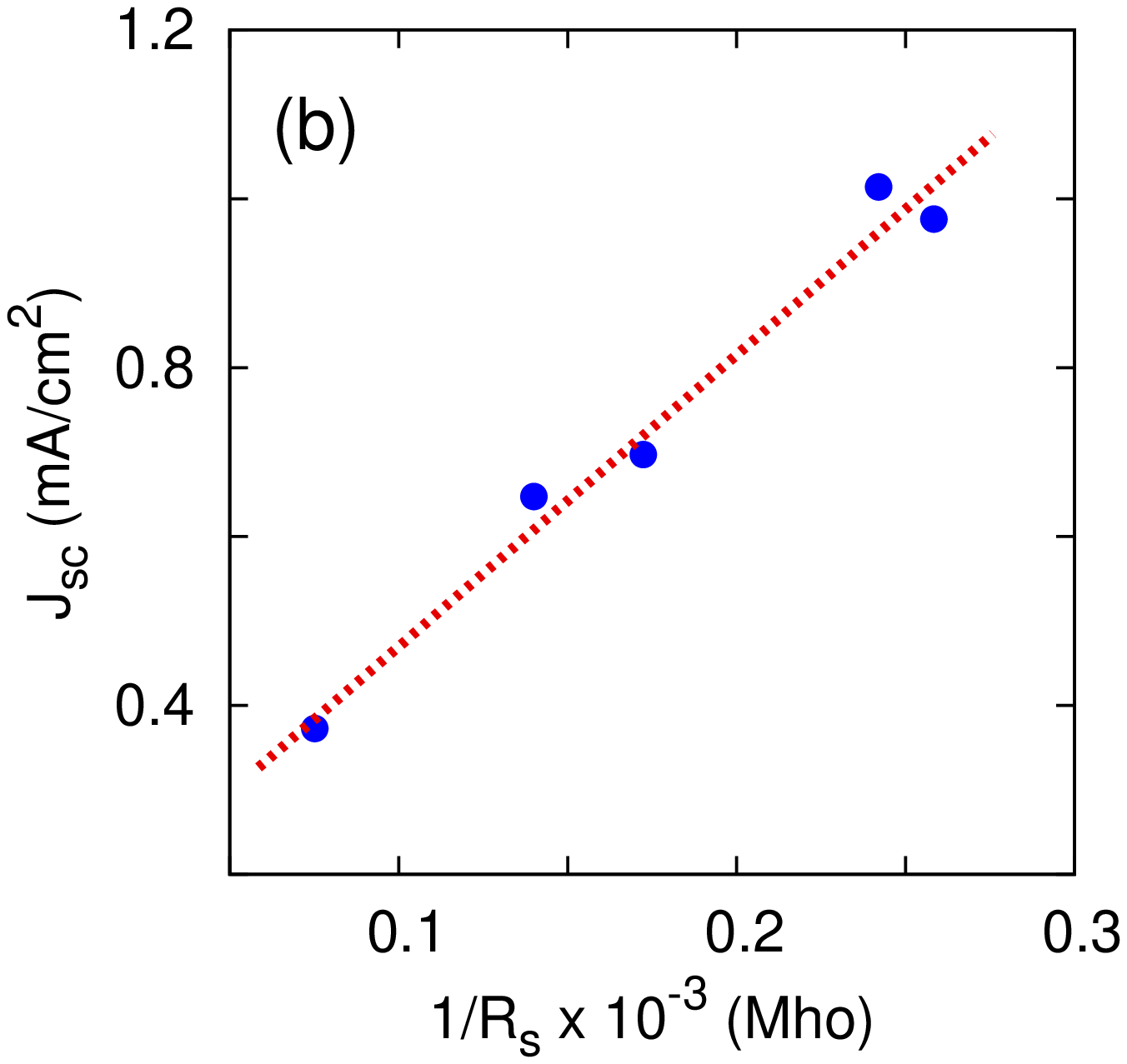, width=2.7in, angle=-90}
\end{center}
\label{voc}
\vskip -0.6cm
\caption{\sl The open circuit voltage (${\rm V_{oc}}$) and the short circuit
current density (${\rm J_{sc}}$) of solar cells as a function of the SnS
grain size within the active layers.} 
\end{figure}

Also we find that the current density (${\rm J_{sc}}$), 
increases with increasing grain size (fig~8b). Considering that we
are talking of the short circuited current and ${\rm R_s << R_p}$, this
current would essentially be flowing through ${\rm R_s}$. As explained
above, the series resistance is essentially due to the active
layer, hence, it would be the prime cause of variation in ${\rm J_{sc}}$. 
Such suggestions, where the short circuit current density is a function of 
the absorbing layer thickness~\cite{david} or grain size \cite{werner,deceglie} 
have been reported before. Hence, to test the argument that the grain size
influences the series resistance which inturn effects ${\rm J_{sc}}$, we
investigated the variation of ${\rm J_{sc}}$ with respect to ${\rm 1/R_s}$.
The prefect linearity confirms that larger grain size results in larger
${\rm J_{sc}}$. This is also conveyed by the data of Table~1.

\begin{figure}[h!]
\begin{center}
\epsfig{file=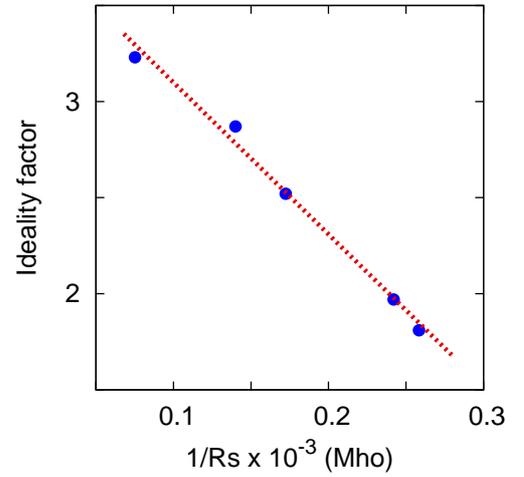, width=2.75in, angle=-90}
\end{center}
\label{nvsgs}
\vskip -0.6cm
\caption{\sl Ideality factor as a function of ${\rm 1/R_s}$. }
\end{figure}

Another important parameter that affects the performance of the solar cell 
is the diode's ideality factor (n). For an ideal solar cell the ideality 
factor is equal to unity~\cite{sze} implying no recombinations take 
place as the charge carriers are moving away from the junction to the
electrodes. A higher `n' implies recombinations are taking place along the
path of the travel. These recombinations maybe taking place at the grain
boundaries or within the grains due to defects. With the I-V characteristics
near linear in the first quadrant, we have used a single diode model to
analyze our device. By not considering the two diode model, we are ruling out 
the
possible contributions from the grain boundaries. The ideality factor for all 
the cells were evaluated at
(V,I)=(0,${\rm I_{sc}}$) (fig~5) using the equation
\begin{eqnarray}
n= {q R_s I_{sc} \over KTln(I_o/ 2I_{sc})}
\end{eqnarray}

\begin{figure}[h!]
\begin{center}
\epsfig{file=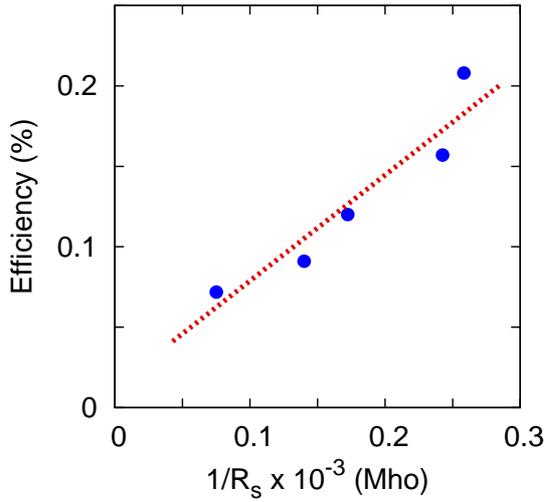, width=3in, angle=-90}
\end{center}
\label{ffvsthick}
\vskip -0.5cm
\caption{\sl Efficiency of the ITO/PEDOT:PSS/SnS/Al structured solar cells 
show a linear dependency on ${\rm 1/R_s}$.
}
\end{figure}

where ${\rm I_o}$ is the reverse saturation current obtained from the dark
I-V characterisitics. The above equation is derived on the assumption 
${\rm R_p>R_s}$. Fig~9 shows that the ideality factor of the cell decreases 
and approaches the value for an ideal cell (${\approx 1}$) \cite{sze} as the 
series resistance decreases or grain size increases. Which means 
that larger the grain size with lower defects results in fewer recombinations 
in the active layer and hence a smaller ideality factor.

Along with the effect of active layer thickness on the solar cell parameters, 
we also studied the effect of light intensity on these parameters.
Charactering parameters of the cells were evaluated for various incident
illumination intensities. Table~2 lists the results for a solar cell with
active layer of 500~nm thickness. Both the series resistance (${\rm R_s}$) 
and shunt resistance (${\rm R_p}$) decrease with the illuminaiton intensity. 
This decrease can be attributed to an increase in conductivity of SnS with 
the increase in the illumination intensity~\cite{illum1,illum2}. This also
explains the increase in short circuit current with increasing illumination.
However, the decrease in ${\rm R_p}$ contributes to a very small increase in 
${\rm V_{oc}}$. With the increase in the intensity, the device's ideality 
factor also approaches that for an ideal solar cell. This is possibly due to 
the increased kinetic energy imparted to the charge carriers that lead to a 
decrease in the recombinations. Though the efficiency of the device does
show an enhancement with increasing intensity, the enhancement is not that 
significant.

As explained in the introduction, research trends on solar cell technology
is concentrated on an effort to improve the efficiency of the solar cell. 
Efficiency
is basically the maximum power generated by the solar cell for the light
intensity flux (or power) incident on it. The results we presented here
are for an ${\rm P_{in}}$ of ${\rm 100~mW/cm^2}$ of Air Mass (AM) 1.5 spectrum.
The maximum power, ${\rm P_{max}}$ was evaluated for the experimental data of
fig~5. The solar cell efficiency was found to vary with grain size. The
variation in efficiency with ${\rm 1/R_s}$ is shown in fig~10. As argued for
fig~8(b), this shows the dependency of efficiency on grain size. Clearly, the 
performance of the solar cell improves with larger 
grain size. Even from Table~1 it is clear that the efficiency of SnS solar
cells increase with grain size, for example, the efficiency varied from ${\rm 
0.02\%}$ \cite{zhi} to ${\rm 1.3\%}$ \cite{reddy} for grain size varing from
10~nm  to 350~nm. However, our structure with 10-20~nm grains in the SnS 
active layer
out-preforms the Wang's \cite{zhi} similar structured cell with same grain 
size. Fact that efficiency increases with grain isze argues for solar cells 
to be made from single crystal wafers. This, however, would push the cost of 
these devices up.
Hence, it would be necessary to study whether plasmonic solar cells of SnS
thin films would be of any practical benefit. As an ending note it is
necessary to point out that the experimental results of our work directly
confirm the intutively indicated ideas of Ghosh et al \cite{ghosh} and those
given theoretically by Guliani et al \cite{renu}.

\begin{table*}
{\bf Table 2:} {\sl Comparison of ITO/PEDOT:PSS/SnS/Al structure under
different illumination intensities.}
\begin{center}
\begin{tabular}{c c c c c c c c}
\hline 
Parameter &&& Intensity (${\rm mW/cm^2}$) &&&&\\
 && 2 & 15 & 30 &&& 100 \\ \hline
\\               

  ${\rm V_{oc}}$ (mV)         && 500 & 505 & 516 &&& 512\\
  ${\rm I_{sc}}$ (${\rm \mu A}$)   && 3.17 & 26.30 & 55.34 &&& 182.61 \\
  ${\rm R_s}$    (${\rm K\Omega}$) && 116.0 & 87.5 & 61.0 &&& 2.94 \\
  ${\rm R_p}$    (${\rm M\Omega}$) && 2.10 & 0.29 & 0.11 &&& 0.05 \\
  ${\rm \eta}$ \%            && 0.13 & 0.14 & 0.14 &&& 0.15 \\
  Ideality factor (n)        && 2.28 & 2.1 & 2.01 &&& 1.97 \\\\ \hline

\end{tabular}
\end{center}  
\end{table*}

\section {Conclusion}
Photovoltaics of ITO/PEDOT:PSS/SnS/Al were fabricated and studied by varying 
the SnS active layer thickness. The varying thickness was found to alter the 
grain size of the nano-crystalline thin films. The fitting to the single
diode model gave parameters that indicated larger grain sized films gave
higher electron-hole generation (${\rm J_{sc}}$) and a lower series
resistance, ${\rm R_s}$. The device's efficiency was also found to be series
resistance dependent.  

This series resistance in turn
depends on the inter-granular distances or distance between the two grains
of the nano-crystalline/ polycrystalline active layer. The incident photons
on the solar cells generate charge carriers that rush to their respective
electrodes. While inter-grain boundaries are sites of recombination which
lead to decrease in the number of charge carriers reaching the electrode (the 
`p' charge carriers reaching the ITO anode), in our case they 
offer a large resistive path. So for easy flow of the charge carriers we
would require a smaller series resistance or in other words
less or no grain boundaries. This would call for a single crystal active
layer. This would clearly mean increased cost of production and doing away
with simple fabrication technique of thin film coating. Hence, we believe future research
would have to focus on methods to tailor materials so as to decrease the series
resistances even for nano/poly-crystalline active layers.

\section*{Acknowledgment}
Authors are thankful to the Department of Science and Technology for funding 
this work under research project SR/NM/NS-28/2010. We also convey our sincere
thanks to Prof. Vinay Gupta, Department of Physics and Astrophysics,
University of Delhi for the help rendered.

\bibliographystyle{modella-num-names}

\end{document}